**Adatom and nanoparticle dynamics on single-atom catalyst substrates**

Matteo Farnesi Camellone[2], Filip Dvořák[1], Mykhailo Vorokhta[1], Andrii Tovt[1], Ivan Khalakhan[1], Viktor Johánek[1], Tomáš Skála[1], Iva Matolínová[1], Stefano Fabris[2,*], and Josef Mysliveček[1,*]

*fabris@iom.cnr.it, josef.myslivecek@mff.cuni.cz

[1]Charles University, Faculty of Mathematics and Physics, Department of Surface and Plasma Science, V Holešovičkách 2, 180 00 Praha 8, Czech Republic

[2]Istituto Officina dei Materiali - CNR-IOM, Area Science Park, Strada Statale 14, km 163.5, 34149 Basovizza – Trieste, Italy

**TOC Graphic**

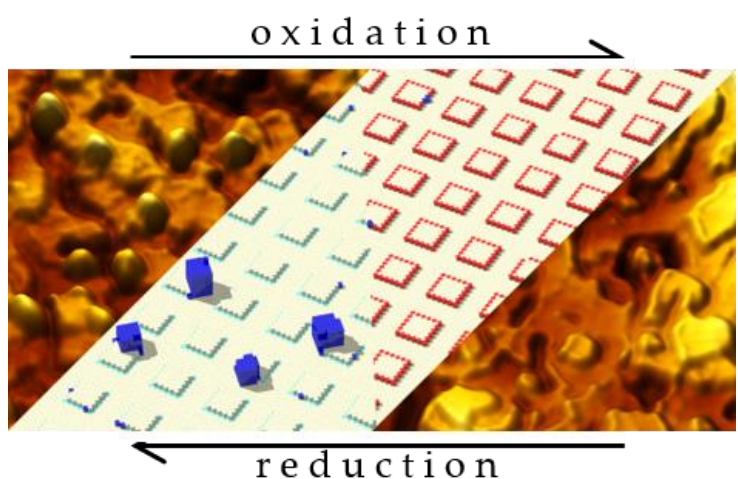

**Abstract**

Single-atom catalysts represent an essential and ever-growing family of heterogeneous catalysts. Recent studies indicate that besides the valuable catalytic properties provided by single-atom active sites, the presence of single-atom sites on the catalyst substrates may significantly influence the population of supported metal nanoparticles coexisting with metal single atoms. Treatment of ceria-based single-atom catalysts in oxidizing or reducing atmospheres was proven to provide precise experimental control of the size of the supported Pt nanoparticles, and, correspondingly, control of catalyst activity and stability. Based on dedicated surface science experiments, ab-initio calculations and kinetic Monte-Carlo simulations we demonstrate that the morphology of Pt nanoparticle population on ceria surface is a result of a competition for Pt atoms between Pt single-atom sites and Pt nanoparticles. In oxidizing atmosphere, Pt single-atom sites provide strong bonding to single Pt atoms and Pt nanoparticles shrink. In reducing atmosphere, Pt single atom sites are depopulated and Pt nanoparticles grow. We formulate a generic model of Pt redispersion and coarsening on ceria substrates. Our model provides a unified atomic-level explanation for a variety of metal nanoparticle dynamic processes observed in single-atom catalysts under stationary or alternating oxidizing/reducing atmospheres, and allows to classify the conditions when nanoparticle ensembles on single-atom catalysts substrates can be stabilized against Ostwald ripening.

**Introduction**

Activity, selectivity and stability of most heterogeneous catalysts depend sensitively on the size of supported metal nanoparticles [1], [2]. Experimental techniques and practical approaches have been developed targeting the nanoparticle size during catalyst activation [3], [4], operation [5], [6] and regeneration [7], [8] with the aim to optimize the catalyst utilization and to improve understanding of nanoparticle dynamic changes throughout the catalyst lifetime. With realization of single-atom catalysts [9], [10] the range of controlled metal nanoparticle sizes has reached a limit of one supported metal atom. Single supported metal atoms do not represent a thermodynamically stable entity on their own. To prevent nucleation of metal atoms to metal nanoparticles, single-atom catalyst substrates must be available, providing sites with strong enough chemical bonding to single metal atoms to overcome metal cohesion [11], [12]. Still, the stabilization of single metal atoms on the catalyst substrate does not guarantee a single-atom catalyst operation. Often, the strong bonding of single atoms to the substrate compromises their catalytic activity [13]–[15], or single atoms are destabilized by interaction with reactants during catalyst operation and metal nanoparticles become the active phase instead [16], [17].

Highly dynamic behavior of metal load on single-atom catalyst substrate has been documented for platinum (Pt) on ceria ($CeO_2$). Ceria can provide strong bonding to metal nanoparticles resulting in high metal dispersion and small nanoparticle size [18]. Recently it was proven that Pt nanoparticles on ceria/alumina substrates are destabilized in oxidizing conditions and at high temperatures and decompose to Pt single atoms bonded to O atoms at surface defects on ceria, particularly at monoatomic surface steps [19]. Such Pt single atoms show small activity towards CO oxidation [20], [21]. Activation of Pt single atoms on ceria for low-temperature CO oxidation can be achieved via treatment of the single-atom catalyst under reducing conditions when in turn, Pt single atom traps are destabilized and three-dimensional Pt nanoparticles of optimal size nucleate [22], [23]. When Pt nanoparticles further grow in size by Ostwald ripening and low-temperature CO oxidation activity is suppressed, catalyst treatment under oxidizing atmosphere can reset the Pt single atom dispersion and make the catalyst ready for new activation [22]. These results indicate exciting possibilities for activation, regeneration and long-time operation of metal nanoparticle catalysts supported on single-atom catalyst substrates [24]–[27] and call for elucidating the mechanisms of Pt redispersion and coarsening on ceria in oxidizing and reducing atmospheres.

We address the issues of Pt redispersion and coarsening in a model system of Pt nanoparticles on $CeO_2(111)$ single crystalline substrate [28]. Flat $CeO_2(111)$ surfaces can be investigated by microscopic and spectroscopic techniques of surface science obtaining direct quantitative information unavailable for real Pt-ceria catalysts: density of surface defects or single atom sites [29], charge state of the surface [30], and occupation of single-atom sites by single Pt atoms [31]. Previously, such quantitative information, combined with ab-initio calculations, allowed identifying Pt single atom dispersion on $CeO_2(111)$ at monoatomic steps acting as single atom traps, configuration of the stable single Pt atom as $Pt^{2+}$ ion bonded to 4 neighboring surface O atoms in a square-planar configuration, and capacity of monoatomic steps on $CeO_2(111)$ to accommodate up to one $Pt^{2+}$ single ion per one step Ce site [31]. Binding energy of Pt in the single-atom sites was investigated as a function of the concentration of Pt and O atoms at $CeO_2(111)$ step edges [32]. Under oxidizing conditions, and in the presence of excess O atoms at step edges, Pt atoms are accommodated at step edges as single $Pt^{2+}$ ions with binding energy $E_1^{ox}$ larger than Pt cohesive energy $E_c$. Under reducing conditions, and in the absence of excess or lattice O atoms at step edges, single Pt atoms are accommodated at step edges as partially charged $Pt^{\delta+}$ species with binding energy $E_1^{re}$ smaller than Pt cohesive energy $E_c$.

In the present work, we perform dedicated surface science experiments revealing the dynamic character of Pt load on $CeO_2(111)$, particularly the thermally activated nature of redispersion of deposited Pt to single $Pt^{2+}$ ions [19], [31]–[33], and Pt coarsening and redispersion in alternating reducing and oxidizing conditions [22], [23], [34]. With further support from ab-initio calculations we formulate a generic model of Pt redispersion and coarsening comprising Pt adatom diffusion on $CeO_2(111)$ terraces, Pt adatom interaction with single-atom traps at $CeO_2(111)$ step edges, and Pt adatom interaction with Pt nanoparticles on $CeO_2(111)$ surface. We show that thermal activation during Pt redispersion and coarsening is required for breaking Pt-Pt bonds in Pt nanoparticles on the way towards minimization of the total energy of $Pt/CeO_2(111)$ system. In oxidizing conditions, capture of Pt adatoms in single atom traps is preferred ($E_1^{ox} > E_c$) and Pt nanoparticles shrink. In reducing conditions, single-atom traps are depopulated ($E_1^{re} < E_c$) and Pt nanoparticles grow by Ostwald ripening. Kinetic Monte Carlo simulations within the proposed model allow us to identify general conditions under which metal nanoparticle catalysts on single-atom catalyst substrates can be optimized and operated in a stable, sinter-resistant manner.

**Results**

**Thermally activated nature of Pt redispersion on $CeO_2(111)$**. To obtain efficient redispersion of Pt deposit to supported Pt single atoms on model and real $CeO_2$ substrates, thermal activation is required [19], [31]–[33]. We investigate this process in detail in a series of surface science ultra-high vacuum (UHV) experiments in which Pt is deposited at 300 K on $CeO_2$(111) samples prepared with well-defined concentrations of surface monoatomic steps providing single-atom Pt adsorption sites [19], [27], [31], [32]. Upon Pt deposition, the samples are subjected to repeated flash annealing to increasingly higher temperatures and characterized with synchrotron radiation photoelectron spectroscopy (SRPES). In the SRPES of Pt 4f, the $Pt^{2+}$ signal corresponds to the amount of occupied Pt single-atom sites, whereas the $Pt^0$ signal represents partially charged $Pt^{\delta+}$ atoms and Pt nucleated as metal nanoparticles ([31]–[33], Figure S1 a, e, i). The fraction of $Pt^{2+}$ signal in the total Pt 4f signal thus allows us to determine the fraction of Pt deposit dispersed as single atoms. No $Pt^{4+}$ signal is observed by SRPES in the present experiments.

We repeat these experiments on four $CeO_2$(111) samples prepared with step densities 6, 11, 15, and 16% monolayer (ML), always depositing 7% ML Pt. 1 ML represents the density of Ce atoms on the $CeO_2$(111) surface ($7.9 \times 10^{14}$ cm$^{-2}$ or 7.9 nm$^{-2}$). The results are summarized in Figure 1. For each sample, a progressive redispersion of Pt is observed with increasing temperature. Three other effects can be identified. At 300 K, the fraction of Pt deposit dispersed as single atoms is increasing with increasing step concentration. Upon final annealing at 700 K, the fraction of Pt deposit dispersed as single atoms adopts two values. For samples where the step density is higher than the amount of deposited Pt, majority of Pt is dispersed as single atoms. For the sample where the step density is lower than the amount of deposited Pt, redispersion of Pt is incomplete. Finally, for the samples showing the almost complete redispersion, an accelerated transient to the redispersed state becomes apparent between 550 and 650 K, indicating the presence of an energy barrier to redispersion.

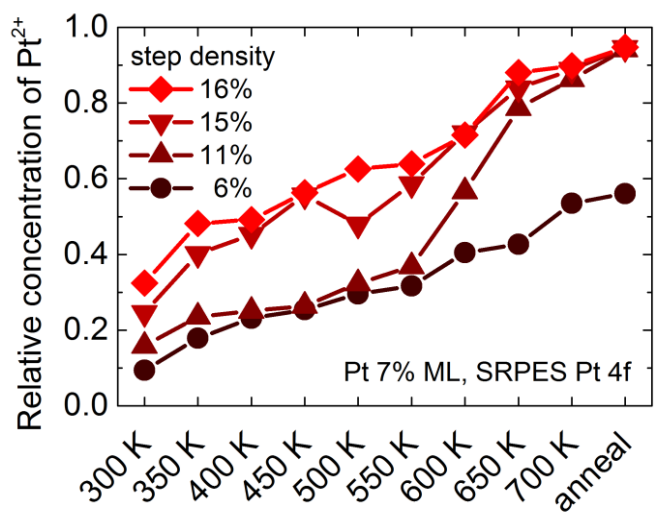

**Figure 1: Redispersion of Pt deposited on CeO$_2$(111) in UHV.** Pt was deposited at 300 K and flash annealed in UHV to increasingly higher temperatures. The final step of the thermal treatment is annealing in UHV at 700 K for 5 min.

**Redispersion and coarsening of Pt load on CeO$_2$(111) in oxidizing and reducing conditions.** Redispersion of Pt load on ceria nanoparticles in oxidizing conditions and subsequent coarsening in reducing conditions have been documented previously by means of transmission electron microscopy (TEM) [19], [22], [35]. Here we perform analogous experiments on model CeO$_2$(111) surfaces by means of SRPES and Scanning Tunneling Microscopy (STM). On the CeO$_2$(111) surface with 15% ML of monoatomic steps we deposit Pt at 300 K, and anneal at 700 K in UHV to obtain Pt redispersion. Afterwards, we perform alternating cycles of sample reduction (in 1×10$^{-5}$ Pa CH$_3$OH) and sample oxidation (in 1×10$^{-5}$ Pa O$_2$, see Experimental and Computational Details). In between the cycles, we characterize the chemical state of the sample by SRPES. We select CH$_3$OH as a reducing agent because, in contrast to CO or H$_2$, it strongly interacts with clean CeO$_2$(111) surfaces and provides a reliable reduction of CeO$_2$(111) surface under UHV conditions of surface science experiments [36]. Reduction with CH$_3$OH does not change the density of monoatomic steps on CeO$_2$(111) surfaces [32].

We repeat these experiments on four $CeO_2(111)$ samples prepared with Pt amounts 2, 6, 16, and 26% ML (0.2 – 2.1 Pt atoms/nm$^2$, cf. Ref. [27]). The results are summarized in Figure 2 a. On all samples, we observe coarsening of the Pt deposit (decrease of the $Pt^{2+}$ signal, increase of $Pt^0$ signal) upon reduction treatments ("red1"–"red5"), and redispersion (increase of the $Pt^{2+}$ signal, decrease of $Pt^0$ signal) upon oxidation treatments ("ox1"–"ox5"). Coarsening and redispersion also influence the total Pt 4f signal due to partial screening of $Pt^0$ signal from metallic Pt nanoparticles. Screening is becoming more efficient for larger nanoparticles; the total Pt 4f signal can thus serve as a qualitative indication of the Pt nanoparticle size. For Pt amounts smaller than the step density (2 and 6% ML Pt) redispersion on the as-prepared samples ("Pt as prep") is almost complete, and subsequent coarsening and redispersion of Pt in the five observed reduction/oxidation cycles is very reversible. For Pt amounts larger than the step density (16 and 26% ML Pt) redispersion on the as-prepared samples is incomplete. Subsequent coarsening and redispersion cycles exhibit a progressive extinction of the $Pt^{2+}$ signal accompanied by a decrease of the total Pt 4f signal indicating an overall growth of metallic Pt nanoparticles at the expense of atomically dispersed Pt.

Parallel to the Pt 4f signal, other SRPES signals were monitored to characterize the chemical state of the samples during reduction and oxidation treatments. Results are shown in Figures S1 and S2. Particularly, reduction of the $CeO_2(111)$ surface was determined by resonant photoelectron spectroscopy (RPES) via so-called Resonant Enhancement Ratio (RER) of the Ce valence band (VB) signal [37]. A complete coarsening (suppression of the $Pt^{2+}$ signal) in the reduction cycles was accomplished by reducing the samples to values of RER > 1.0 [32]. Upon oxidizing cycles, RER decreases to < 0.1 corresponding to almost fully stoichiometric $CeO_2(111)$ [32]. Reduced $CeO_{2-x}(111)$ surface is highly reactive towards water and alcohols resulting in strong OH adsorption at the surface (Figure S1 g, h). The use of organic reductant (CH$_3$OH) in the reduction cycles results in the presence of carbonaceous adsorbates at the surface upon both reducing and oxidizing cycles (Figure S1 d, h, l). This indicates that the redispersion provided to Pt by ceria-based single-atom catalyst supports is robust with respect to organic and carbonaceous contaminants.

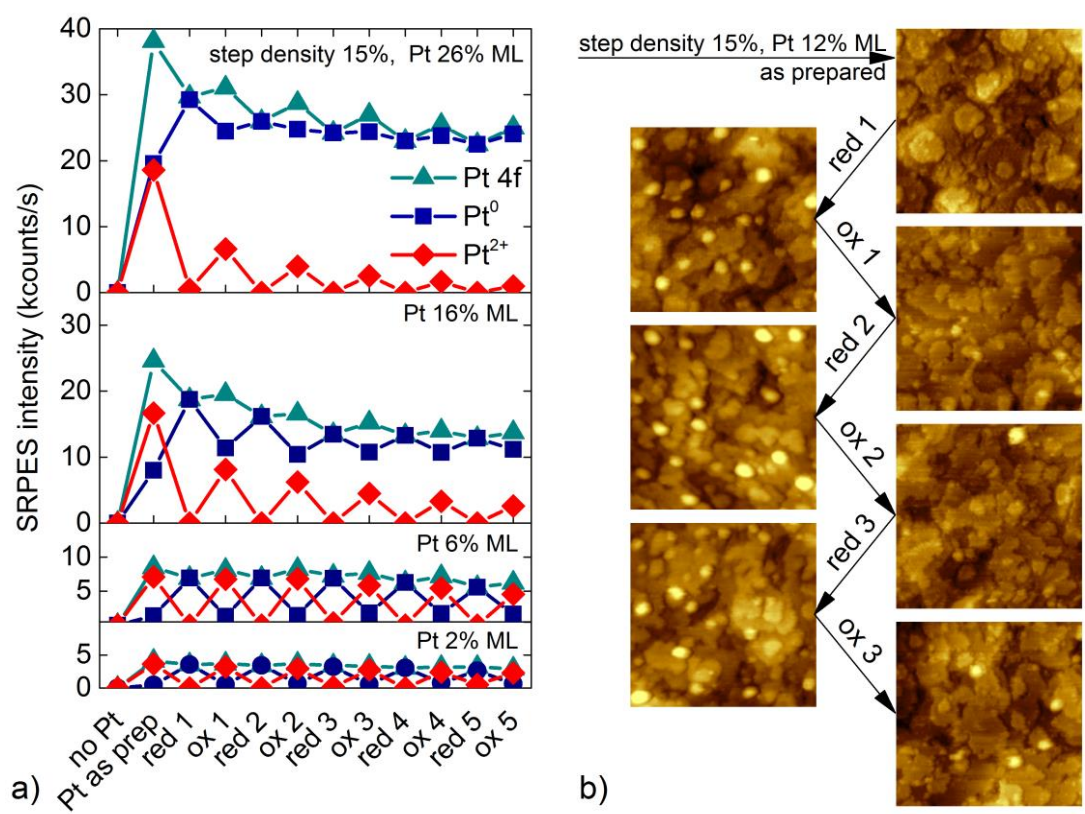

**Figure 2: Redispersion and coarsening of Pt load on CeO$_2$(111) in oxidizing/reducing conditions.** Pt was deposited at 300 K and annealed at 700 K in UHV ("Pt as prep") followed by annealing at 600 K in alternating reducing (CH$_3$OH, "red") and oxidizing (O$_2$, "ox") conditions. Experiments were performed for Pt amounts between 2 and 26% ML deposited on CeO$_2$(111) substrates with 15% ML monoatomic steps. a) Pt$^{2+}$ (single atoms), Pt$^0$ (metallic clusters), and total Pt 4f signal in SRPES spectrum. b) STM images for 12% ML Pt. Occupied state images, 45 × 45 nm$^2$, sample voltage 7.5 V, tunneling current 45 pA.

An STM experiment illustrating the morphology of the model Pt/CeO$_2$(111) samples was performed for 12% ML Pt (0.9 Pt atoms/nm$^2$, cf. Ref. [27]) deposited on the CeO$_2$(111) surface with 15% ML steps. The results are summarized in Figure 2 b. On the as-prepared samples with fully redispersed Pt, only (111) terraces and monolayer-high steps of CeO$_2$(111) surface are apparent. Upon reduction treatment, nanoparticles decorating ceria step edges appear. The observed nanoparticle density corresponds to an average nanoparticle size of 120 Pt atoms. Upon oxidation treatment, the nanoparticles disappear, however, the efficacy of this process seems to be decreasing for the consecutive reduction/oxidation cycles. Pt nanoparticles on the partially reduced CeO$_2$(111) surfaces were best viewed by STM in occupied states with tunneling voltage increased to 7–8 V. In empty states, STM contrast of the Pt nanoparticles was lower and comparable to the contrast of the OH terminated, oxygen-deficient CeO$_2$(111) (Figure S3). STM experiments also confirm that upon both oxidation and reduction treatments morphology and step density of CeO$_2$(111) substrates remain unchanged.

The experiments summarized in Figures 1 and 2 reveal a variety of dynamic processes related to redispersion and coarsening of the Pt load on CeO$_2$(111). A common property of all observed processes is their thermally activated nature. Performing the experiments on flat model CeO$_2$(111) substrates using surface science methods allowed quantification of the density of available single-atom sites. Many observed processes are influenced by the capacity of the CeO$_2$(111) surface to accommodate Pt in single-atom traps, i.e. by the ratio of the density of the monoatomic steps on CeO$_2$(111) to the amount of deposited Pt. Some general phenomena can be identified, e.g. coarsening of Pt nanoparticle populations for the cases when metallic Pt nanoparticles prevail.

**Activation barriers for filling Pt single atom traps on CeO$_2$(111).** Understanding the thermally activated nature of the Pt redispersion on CeO$_2$ in oxidizing conditions requires considering Pt transport processes on CeO$_2$(111) surface at the atomic scale. Pt nanoparticle diffusion can be excluded, because single atom traps at CeO$_2$(111) step edges must be populated by diffusing species containing single Pt atoms. In some studies, Pt mass transport through gas phase via volatile PtO$_2$ is invoked [27], [38], [39]. On geometrically flat model CeO$_2$(111) substrates, however, formation and desorption of PtO$_2$ would lead to a fast removal of Pt deposit which is not observed. Ab-initio calculations predict that due to small binding energy of a PtO$_2$ molecule on CeO$_2$(111), surface diffusion of PtO$_2$ does not take place [27]. The presence of carbonaceous species on our samples (Figure S1 d, h, l) may invoke consideration of Pt transport via Pt carbonyls [40]. This mechanism however cannot account for all observed phenomena, e.g. for the Pt redispersion after deposition (Figure 1, Figure S1 d). Moreover, monodispersed Pt$^{2+}$ ions at ceria step edges do not interact with CO [31]. It seems reasonable to assume mass transport of Pt on CeO$_2$(111) surfaces via diffusion of Pt single atoms — adatoms. Theoretical studies predict the binding energy of Pt adatom on CeO$_2$(111) surface 3.3 eV [31], and the activation energy for Pt adatom hopping on clean CeO$_2$(111) surface 0.5 eV, rendering surface diffusion of Pt adatoms a fast and efficient process [41].

Single-atom traps for Pt adatoms are localized at the step edges of CeO$_2$(111) [19], [27], [31], [32] and must be occupied by surface diffusion of Pt adatoms from upper or lower CeO$_2$(111) terraces adjacent to the step edges. This process may account for the experimentally observed thermal activation of Pt redispersion provided that energy barriers preventing adatom attachment at the step edge from the upper or the lower terrace are present [42], [43]. Here we estimate such step-edge barriers for Pt adatom capture at $CeO_2(111)$ monoatomic step edges based on density functional theory (DFT) calculations [44].

Following our previous works on Pt incorporation and reactivity at $CeO_2(111)$ step-edges [31], [32], [45], we select two different low-energy step geometries, which we label as Type I and Type II [46]. For each step geometry, we consider both stoichiometric step edges as well as step edges in the presence of excess oxygen atoms decorating the steps (Figure S4). We have previously demonstrated that excess O atoms bind preferentially at step edges leading to the formation of stable peroxide species. Peroxide species in turn determine the preferential binding of Pt adatoms to the oxidized step edges, and increase the Pt binding energy by 1.6 eV with respect to stoichiometric step edges. This large increase in Pt binding energy correlates with the formation of characteristic square-planar $PtO_4$ units [31]. The adsorption energy of Pt on all the considered model steps is summarized in Table S1.

In order to quantify the step-edge barrier for the Pt incorporation at $CeO_2(111)$ step edges via surface diffusion, we have calculated the minimum energy path corresponding to the diffusion of a Pt adatom from a $CeO_2(111)$ terrace to a step edge. This process has been calculated considering diffusion paths from both sides of the step edge (i.e. from the upper and lower terraces), for both model step edges (Type I and Type II), and for different degrees of step-edge oxidation (stoichiometric and with excess oxygen). The minimum energy paths have been computed using the climbing image nudged elastic band (CI-NEB) method, which provided the activation energy for Pt diffusion at the ceria step edge. Schematic illustration of the calculation for Step Type II with excess oxygen is in Figure 3, all calculated configurations are summarized in Figure S5.

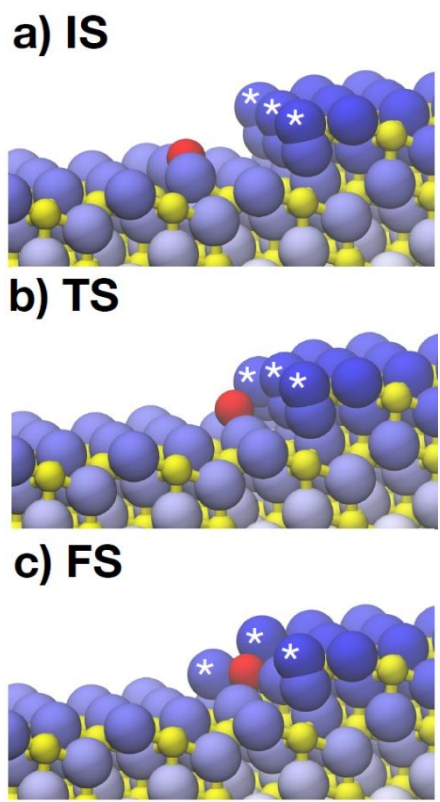

**Figure 3. Representative configurations of the calculated minimum energy path for the Pt adatom diffusion and incorporation at Type II Excess Oxygen step edge.** Panels a-c displays the initial (IS), transition (TS) and final (FS) states of the process. Pt atom red, Ce atoms yellow, O atoms blue. Excess O atoms are indicated with *.

|  | From lower terrace | From upper terrace |
| --- | --- | --- |
| **Step Type I Stoichiometric** | 0.47 eV | 0.75 eV |
| **Step Type II Stoichiometric** | 0.59 eV | 0.42 eV |
| **Step Type I Excess Oxygen** | 0.60 eV | 0.60 eV |
| **Step Type II Excess Oxygen** | 0.16 eV | 0.40 eV |

**Table 1: DFT calculations of activation energies for Pt adatom capture in single-atom traps on CeO$_2$(111).**

The resulting step-edge barriers are reported in Table 1. We compare the step edge barriers to the activation energy $E_d$ for Pt diffusion on the stoichiometric $CeO_2(111)$ terrace, $E_d = 0.50$ eV. For most cases, the activation energies for Pt adatom capture at $CeO_2(111)$ single-atom traps differ from the activation energy of Pt adatom diffusion on the (111) terrace by 0.1 eV or less, where 0.1 eV is the estimated error of the calculation. In one case (Step Type II excess oxygen/lower terrace), Pt diffusion towards the step edge is enhanced, and in one case (Step Type I stoichiometric/upper terrace), the diffusion requires additional activation. Still, Steps Type I stoichiometric can be easily occupied from the lower terrace. Overall, we can conclude that the DFT calculations do not predict significant activation barriers for Pt adatom capture in single-atom traps on $CeO_2(111)$.

**Generic model of metal redispersion and coarsening on single-atom catalyst substrates.**

The absence of the step-edge barrier for Pt adatom capture in single-atom traps on $CeO_2(111)$ step edges resulting from our DFT calculations (Figure 3, Table 1) indicates that Pt adatom capture is a diffusion-limited process and the thermal activation in the present (Figures 1, 2) and previous experimental observations [19], [24], [31]–[33] is rather required to increase the Pt adatom concentration on the samples, and thus Pt adatom availability for a capture. The only source of Pt adatoms on our samples can be Pt nanoparticles. In such case, Pt distribution on the sample surface must be a result of competition between adatom capture in single-atom traps and adatom capture by Pt nanoparticles.

For the formulation of a model of Pt redispersion and coarsening on $CeO_2(111)$ a suitable description of Pt cohesion in Pt nanoparticles is required. While ab-initio calculations of stability of supported metal clusters and ab-initio calculations of atom detachment from the clusters are available [47]–[49], they are restricted to specific cluster geometries and cannot effectively describe the broad range of cluster configurations at elevated temperatures. We aim at formulating a generic model, i.e. a model replicating the observed Pt redispersion and coarsening phenomena based on a minimum amount of assumptions. In such case, it is favorable to describe Pt cohesion within the frame of a bond-counting model, when the energy cost of Pt adatom detachment from a Pt nanoparticle is considered proportional to the number of broken nearest-neighbor Pt-Pt bonds [50], [51]. Kinetic Monte Carlo implementations of bond counting models correctly replicate nucleation and Ostwald ripening of metal nanoparticles [52] and can be modified for considering additional kinetic phenomena, e.g. in anisotropic or multicomponent systems [53], [54].

To describe Pt redispersion and coarsening on ceria, we perform kinetic Monte Carlo (KMC) simulations within a bond counting model on a plain cubic lattice. On the plain cubic lattice, Pt cohesive energy is equal to three times the energy of the broken Pt-Pt bond, $E_c=3\times E_n$ [55]. Ceria substrate is considered immobile and containing single-atom steps. Steps type I and II on $CeO_2(111)$ are represented by {110} steps on the plain cubic lattice. {110} steps correctly replicate the capacity of steps Type I and II to accommodate up to 1 Pt single atom per 1 step site without creating Pt-Pt bonds, see Figures S6, S7. Binding energy of the Pt atom at step edge site can be switched between that corresponding to oxidized step edge $E_1^{ox}$ (Figure S6 a) and reduced step edge $E_1^{re}$ (Figure S6 b). All Pt adatom hops in the model are assigned the same activation energy $E_d$, and no additional energy barrier is associated with Pt attachment or detachment to and from the step edges and Pt nanoparticles. Energy parameters of the model are selected in the way that they obey the relationship $E_1^{ox} > E_c > E_1^{re}$ determined previously based on ab-initio calculations [31], [32]. In order to achieve manageable computational times, $E_1^{ox}$, $E_c$, and $E_1^{re}$ are set approximately 1 eV lower than predicted by ab-initio calculations resulting in an effective downshift of the simulation temperatures by 100-150 K relative to the temperatures in the experiment. Parameters of the KMC simulation are summarized in Table 2 while the relation of binding energies in the KMC simulation to the results of ab-initio calculations is illustrated in Figure S8.

| activation energy | | |
|---|---|---|
| Pt adatom diffusion | $E_d$ | 0.5 eV |
| **binding energy** | | |
| Pt–Pt nearest-neighbor | $E_n$ | 0.35 eV |
| Pt cohesive | $E_c$ | 1.05 eV |
| Pt at oxidized step | $E_1^{ox}$ | 1.3 eV |
| Pt at reduced step | $E_1^{re}$ | 0.7 eV |

**Table 2: Activation and binding energies of Pt adatom in the KMC simulations.** Binding energies are relative to the binding energy of Pt adatom on the oxide surface.

KMC simulations performed within the proposed generic model qualitatively replicate and allow to interpret the experimental data. For the experiment on thermal activation of Pt redispersion presented in Figure 1, we perform KMC simulation depositing 5% ML Pt at 300 K on substrates containing 6%, 10% or 17% monoatomic steps. Ceria substrates before Pt deposition are considered partly oxidized with oxidized and reduced steps each representing 50% of the step-edge sites (Figure S6 c). Upon Pt deposition, we simulate 30 s annealing cycles at temperatures 350 K, 400 K … 700 K and evaluate the Pt dispersion on the samples at the end of each annealing cycle. In analogy to previous experimental and theoretical studies [31]–[33], the intensity of the simulated $Pt^{2+}$ SRPES signal is considered proportional to the amount of isolated Pt adatoms at oxidized steps. All other Pt atoms in the simulation contribute to the intensity of the simulated $Pt^0$ SRPES signal (Figure S7). For the evaluation of the simulated $Pt^0$ SRPES signal, the exponential attenuation of the signal from subsurface Pt atoms is taken into account (Figure S7).

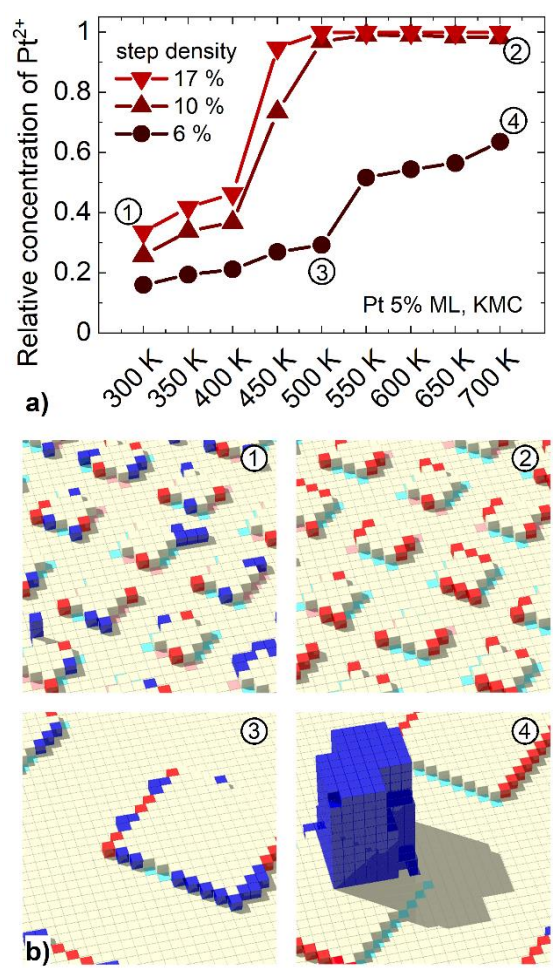

**Figure 4: KMC simulation of Pt redispersion in single-atom traps.** a) Pt deposition at 300 K and intermittent annealing to increasingly higher temperatures. b) KMC morphologies corresponding to the points ①-④ in a). Color coding: red - $Pt^{2+}$ ions, blue - metallic Pt, pink - oxidized step, cyan - reduced step.

Results of the simulation are summarized in Figure 4. At 300 K, $E_d$ of Pt on ceria (0.5 eV) is sufficient for all deposited atoms to reach step edges. With increasing step density and decreasing distance between the steps the probability for the adatom to reach oxidized step is increasing, resulting in the increasing fraction of Pt atoms dispersed as $Pt^{2+}$ ions. Other deposited adatoms reach reduced steps or nucleate small metallic Pt aggregates at step edges (Figure 4 b, panels ① and ③). The resulting morphology at 300 K is kinetically limited and subsequent annealing at progressively higher temperatures brings the system to more energetically favorable configurations increasing the occupancy of oxidized steps. Upon annealing at 700 K, the dispersion of Pt as $Pt^{2+}$ ions is complete for the samples with density of oxidized steps exceeding the amount of deposited Pt (Figure 4 b, panel ②), and maximized for the sample with density of oxidized steps lower than the amount of deposited Pt (Figure 4 b, panel ④). At temperatures between 400 K and 550 K an accelerated transient to the $Pt^{2+}$ state is observed that can be attributed to thermal activation of characteristic system configurations, particularly Pt atoms with 2 nearest neighbors, and Pt atoms in shallow single-atom traps (cf. Figure 4 b, panel ③).

KMC simulations clearly illustrate that the thermal activation required for redispersion of Pt load on $CeO_2(111)$ is due to kinetic limitations preventing the Pt adatoms from reaching oxidized step edges rather than due to the microscopic nature of the interactions between Pt adatoms and steps (cf. Table 1). Assuming sufficient thermal activation, we now attempt at simulating Pt coarsening and redispersion in reducing and oxidizing conditions (Figure 2). On $CeO_2(111)$, changes in the oxygen content at monolayer step edges result in the changes of binding energy of Pt single-atoms — from low $E_1^{re}$ at oxygen deficiency to high $E_1^{ox}$ at oxygen surplus and vice versa (Table S1, Table 2, [31], [32]).

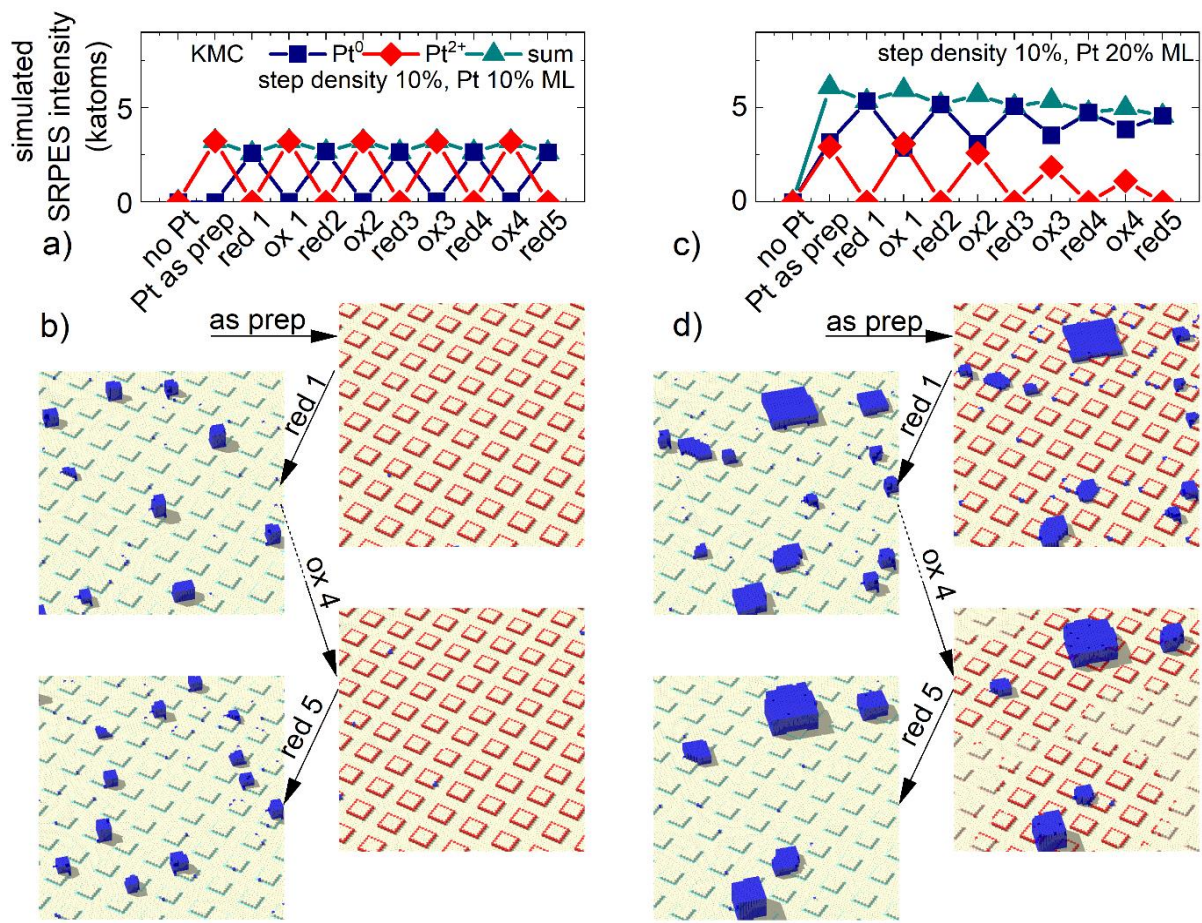

**Figure 5: KMC simulation of redispersion and coarsening of Pt load.** Redispersion and coarsening are induced by alternatingly switching the step edges from oxidized to reduced. Step density 10% ML, Pt amount (a), (b) 10% ML, (c), (d) 20% ML. (a), (c) Single atom ($Pt^{2+}$), metallic ($Pt^0$), and Pt total ("sum") simulated SRPES signal. (b), (d) KMC morphologies. Color coding: red - $Pt^{2+}$ ions, blue - metallic Pt, pink - oxidized step, cyan - reduced step.

Results of the KMC simulation are summarized in Figure 5. Simulations were performed on substrates containing 10% of monolayer high steps and Pt deposit amount 10% ML (Figure 5 a, b) and 20% ML (Figure 5 c, d). Pt was deposited at 300 K and redispersed at temperature 550 K for 200 s under oxidizing conditions (all steps oxidized, Figure S6 b). Afterwards, the temperature was kept at 550 K and the substrate condition was alternated in 200 s time intervals between reduced (Figure S6 a) and oxidized (Figure S6 b). At the end of each time interval, simulated Pt signals ($Pt^{2+}$, $Pt^0$) and system morphology were evaluated.

In both calculations, we observe qualitative replication of the dynamic phenomena identified in the experiment (Figure 2 a): coarsening of the Pt deposit (decrease of the $Pt^{2+}$ signal, increase of $Pt^0$ signal) upon reduction treatments ("red1"–"red5"), redispersion (increase of the $Pt^{2+}$ signal, decrease of $Pt^0$ signal) upon oxidation treatments ("ox1"–"ox4"), and corresponding changes of the total Pt 4f signal due to the screening of $Pt^0$ signal from 3D Pt nanoparticles. For Pt amount comparable to the step density (10% ML Pt) Pt redispersion on the as-prepared samples ("Pt as prep") is almost complete, and subsequent coarsening is very reversible. For Pt amount larger than the step density (20% Pt) Pt redispersion on the as-prepared samples is incomplete and subsequent coarsening and redispersion cycles exhibit progressive growth of metallic Pt nanoparticles at the expense of atomically dispersed Pt. Simulated morphologies correspond to the morphology observed in STM (Figure 2 b) and provide a real space illustration of the observed dynamic phenomena.

**Stable and unstable nanoparticle distributions within the generic model.**

Obtaining qualitative agreement between experimental and simulated data on adatom and nanoparticle dynamics allows us to investigate conditions for stable and unstable nanoparticle populations within the proposed generic model. At stationary oxidizing conditions, the parameter determining stability or instability of the resulting Pt population is the density of the oxidized steps $\rho_{ox}$. A trivial stable solution is obtained for Pt amount $\rho_{Pt} \leq \rho_{ox}$ when, ultimately, all Pt is redispersed as single atoms. A nontrivial stable solution is obtained for Pt amount slightly exceeding $\rho_{ox}$, $\rho_{Pt} = \rho_{ox} + \varepsilon$. This solution is illustrated in Figure 6 a. We simulate redispersion of Pt clusters of average size ~80 atoms under oxidizing conditions and at 550 K. Redispersion is complete after 400 s. Afterwards, Pt population remains stable until 3600 s. Pt atoms exceeding the capacity of oxidized steps remain highly dispersed, forming subcritical 2D $Pt^0$ clusters with a prevailing size of 3 Pt atoms, see Figure 6 b.

In the present discussion, and in Figures 6 and 7, we evaluate the density of the subcritical 2D $Pt^0$ clusters separately from the density and average size of 3D $Pt^0$ clusters. Subcritical clusters originate from attachment of Pt adatoms to oxidized steps decorated by $Pt^{2+}$, and, given the density of subcritical clusters, size distribution of metallic Pt nanoparticles in the model (3D clusters + 2D subcritical clusters) must be considered bimodal. The existence, morphology and metallic character of subcritical clusters are rationalized by previous DFT studies of Pt structures on ceria step edges [31], [32]. Density of subcritical clusters on the surface reflects in part the intensity of Pt surface mass transport, cf. the correlated character of the fluctuations of the average nanoparticle size and the density of subcritical clusters in Figure 6 a. Subcritical clusters can be generally considered to provide different chemical reactivity compared to Pt single atoms in deep traps or to 3D Pt nanoparticles [26].

Unstable nanoparticle distributions at stationary oxidizing conditions are obtained for $\rho_{Pt} > \rho_{ox} + \varepsilon$. The situation is illustrated in Figure 6 c, d. We simulate redispersion of Pt nanoparticles of average size ~500 atoms under oxidizing conditions and at 550 K. After 400 s $Pt^{2+}$ population at oxidized steps is maximized. Redispersion is however incomplete since Pt exceeding the capacity of single-atom traps remains on the surface in the form of 3D clusters (Figure 6 d). 3D clusters undergo Ostwald ripening and cause the unstable character of the Pt population with nanoparticle density and mean nanoparticle size changing at times of 3600 s (Figure 6 c) and beyond.

At stationary reducing conditions, only an unstable solution is available. 3D clusters undergo Ostwald ripening which is, in the absence of additional constraints on characteristic system sizes [56], [57], a nonstationary process [58].

In alternating oxidizing/reducing conditions, the stability of the nanoparticle populations requires, in addition to sufficient $\rho_{ox}$, a suitable timing. We perform KMC simulations of the nanoparticle population for $\rho_{Pt}$ = 10% ML, and $\rho_{ox}$ =17% ML ($\rho_{Pt} < \rho_{ox}$), and at simulation temperature 650 K when the redispersion and coarsening times are reduced to an order of single seconds, comparable to, e.g. alternating oxidizing/reducing conditions in three-way automotive catalytic converters [59], [60]. Oxidizing and reducing conditions are simulated as in calculations presented in Figure 5 – all step-edge traps are set oxidized (Figure S6 b) or reduced (Figure S6 a).

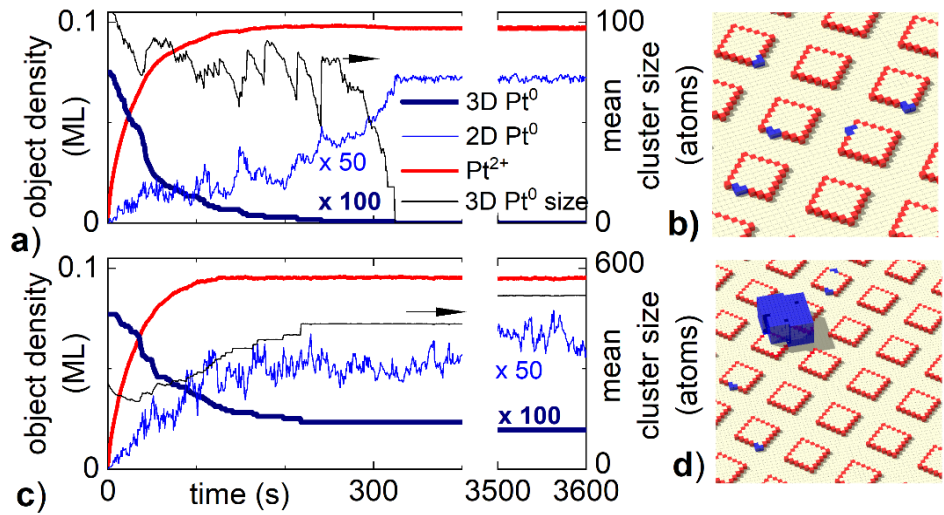

**Figure 6: KMC simulations of stable (a, b) and unstable (c, d) nanoparticle populations upon redispersion of 3D Pt clusters at stationary oxidizing conditions.** Step density 10% ML, Pt amount (a, b) (10+ε) % ML, ε=0.15 % ML, (c, d) 15 % ML. (a, c) time evolution of density of $Pt^{2+}$ single atoms, subcritical 2D $Pt^0$ clusters, 3D $Pt^0$ clusters, and average 3D $Pt^0$ cluster size. (b, d) KMC morphologies at 3600 s.

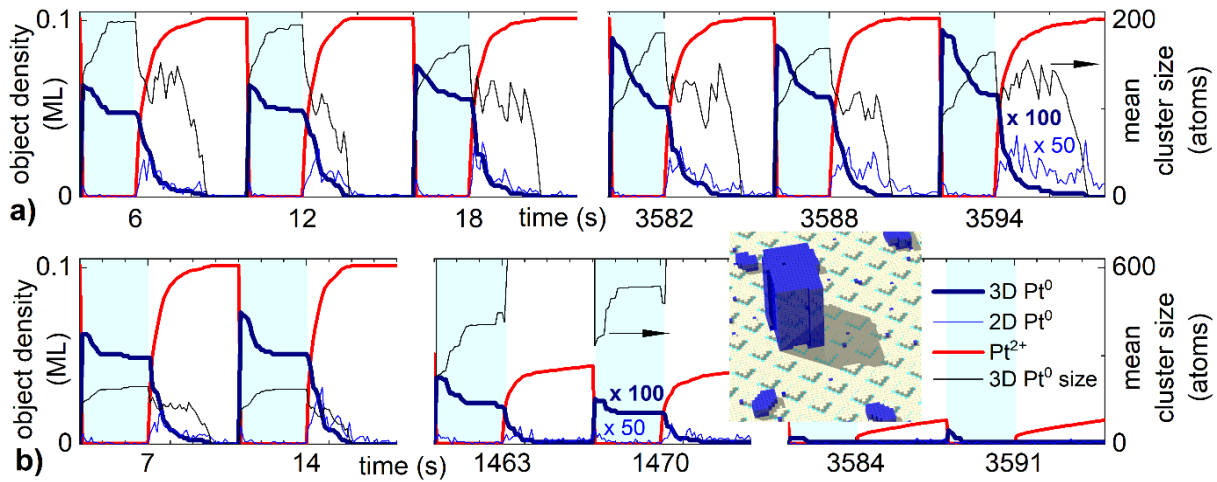

**Figure 7: KMC simulations of stable (a) and unstable (b) nanoparticle populations during alternating oxidizing/reducing conditions.** Step density 17% ML, Pt amount 10% ML. Cycle times: (a) 2 s reduction, 4 s oxidation (b) 3 s reduction, 4 s oxidation. Reduction cycles are highlighted light blue. Inset in (b): KMC morphology at 1470 s.

Stable nanoparticle population can be obtained for alternating 2 s reduction cycles and 4 s oxidation cycles as illustrated in Figure 7 a for simulation times up to 3600 s. Within each oxidation cycle, a complete redispersion of Pt deposit at oxidized steps is obtained. Within reduction cycles, Pt nanoparticles are formed via nucleation and coarsening of Pt adatoms released upon the change from oxidized steps to reduced steps.

Increasing the duration of the reduction cycles to 3 s while keeping the oxidation cycles 4 s leads to unstable nanoparticle population illustrated in Figure 7 b. The breakdown of the stable nanoparticle population is triggered by an incomplete redispersion of Pt in one particular oxidation cycle. This has a consequence of Pt nanoparticles remaining on the surface at the beginning of the next reduction cycle (cf. Figure 5 d). Pt adatoms released upon the change from oxidized steps to reduced steps are then preferentially captured at the remaining Pt nanoparticles resulting in their fast coarsening (see the inset of Figure 7 b), and the breakdown of the nanoparticle population, within the spatial limits of KMC simulations, to the fastest growing single Pt particle. Once this process is started, it cannot be reverted by the intermittent 4 s oxidation cycles. For restoring the complete redispersion, oxidation/reduction cycling must be interrupted and a sufficiently long treatment in oxidizing conditions applied.

**Discussion**

The presented KMC simulations within the generic model of Pt adatom and nanoparticle dynamics on single-atom catalyst surface allow us to conclude that in order to obtain a stable nanoparticle population in stationary oxidizing or alternating reducing/oxidizing conditions, complete Pt redispersion during oxidation must be reached. For complete redispersion, conditions of both sufficient density of step edges allowing the accommodation of the majority of deposited Pt, and sufficient time allowing redispersion of Pt clusters of a particular size must be fulfilled. KMC simulations allowed identifying types of Pt objects constituting Pt populations on single-atom catalyst substrates. Besides Pt single atoms and 3D Pt clusters, subcritical 2D Pt clusters should be considered. Presented examples illustrate that within the generic model, oxidation and reduction treatments can be configured for optimizing the size, surface concentration and stability of any of the considered Pt objects.

The presented view of Pt redispersion and coarsening on ceria as a competition for Pt adatoms between single-atom traps at ceria step edges and Pt nanoparticles can be, in many cases, applied to real ceria catalysts. Monoatomic step edges are ubiquitous on well-annealed nanoparticulate and nanoporous ceria supports [19], [27], [61], [62]. Excess O atoms at ceria step edges required for strong bonding of single $Pt^{2+}$ ions are present already in the conditions of model UHV experiments [31], [32]. Excess O in the form of peroxo and superoxo species is readily observed and available for $Pt^{2+}$ stabilization also on real ceria catalysts in air or in lean, oxygen-containing reaction mixtures [63]–[65]. As is the case of model experiments, monoatomic step edges on real Pt/ceria catalysts exhibit limited capacity to accommodate single $Pt^{2+}$ species during high-temperature treatment in air. Full Pt dispersion to $Pt^{2+}$ species was observed for $\rho_{ox} > \rho_{Pt}$. For $\rho_{ox} < \rho_{Pt}$ saturation of step edges was observed, resulting in coexistence of isolated $Pt^{2+}$ species and Pt nanoparticles ([27], cf. Figure 5). In real Pt/ceria catalysts, mass transport of Pt on the ceria surface via Pt adatom diffusion can be efficient at lower temperatures, and within the surface of individual ceria nanoparticles as observed e.g. by TEM in [22]. At higher temperatures, or for Pt transport among the ceria nanoparticles sublimation of $PtO_2$ species and Pt mass transport through gas phase may be invoked [27], [38]. The efficacy of mass transport via $PtO_2$ in real catalysts may however need verification by further experiments, e.g. to determine its substrate dependence when Pt load seems unstable on alumina [66] and stable on ceria [57], [59].

The change from oxidizing conditions to reducing conditions in the presented model is considered instantaneous (Figure S6). In real catalysts, a chemical reaction in oxygen deficient rich reducing reaction mixture must proceed removing O atoms (excess O or lattice O) from $Pt^{2+}$ complexes at step edges. This reaction is expected to proceed fastest at the $Pt^{2+}$ complexes, and at superoxo, peroxo, and lattice O sites at ceria step edges since the binding energies of O at these sites are smaller than on the facets of ceria nanoparticles [31], [67]. The amount of removed oxygen atoms required to deactivate all $Pt^{2+}$ traps at ceria step edges and initiate nucleation and growth of Pt nanoparticles is on the order of monoatomic step density, i.e. 10 % of surface Ce atoms [32]. With bulk sensitive experimental methods employed for characterization of Pt/ceria catalysts under reaction conditions such amount of removed O or reduced $Ce^{3+}$ may be difficult to detect and the deactivation of $Pt^{2+}$ traps in reducing conditions can happen before the reduction of ceria bulk is observed [23].

Apart from single Pt atoms in $Pt^{2+}$ chemical state and metallic Pt nanoparticles some studies on real Pt/ceria catalysts report one dimensional (1D) and two-dimensional (2D) Pt objects of oxidized or mixed oxidized/metallic chemical state and a specific catalytic activity different from single $Pt^{2+}$ ions or metallic Pt nanoparticles [26]. Our model suggests that such objects may be related to ceria step edges densely populated with $Pt^{2+}$ ions or to 2D Pt objects at step edges (Figure 6b). 2D decoration of ceria step edges with Pt in a mixed oxidized/metallic chemical state has been proposed earlier based on ab-initio calculations [32].

Generic model of Pt redispersion and coarsening introduced in this work defines and considers a minimal set of kinetic processes required for qualitative replication of nanoparticle dynamics phenomena on single-atom catalyst substrates and for a general classification of stability of nanoparticle population under stationary or alternating oxidation and reduction conditions. For future quantitative predictions of the evolution of supported nanoparticle size in real catalysts and reaction mixtures, further details of the involved kinetic processes must be clarified at atomic scale. A prerequisite would be the implementation of the KMC simulation on a FCC lattice including kinetic processes relevant to FCC Pt nanoparticle equilibration [55], [68]. In the present model, the interaction of the catalyst with oxidizing or reducing atmospheres is influencing only the properties of the single-atom traps. In reality, Pt mass transport across the ceria surface, nucleation of Pt nanoparticles, and the rate of Pt nanoparticle growth or disintegration may vary substantially as a function of concentration of O vacancies, reactants, intermediates or products on ceria and Pt nanoparticle surfaces [69]–[71]. This is corroborated by recent experiments on real Pt/ceria catalysts that reveal a strong influence of the composition of rich reaction mixtures on the activation of Pt/ceria catalysts by redispersion [22], [23]. Sensitivity of the redispersion and coarsening processes to the composition of reaction mixtures provides a broad range of topics for further dedicated studies of adatom and nanoparticle dynamics on single-atom catalyst substrates.

**Conclusions**

Based on surface science studies and DFT analysis of Pt adatom and nanoparticle dynamics on $CeO_2(111)$ surface, we formulate a generic model of metal redispersion and coarsening on single-atom catalyst substrates. The model comprises fast metal adatom diffusion and adatom interaction with single-atom traps or metal nanoparticles on the catalyst substrate. Activation barriers inherent to Pt redispersion and coarsening on ceria-based single-atom catalyst substrates are associated with transitions of the morphology of supported Pt to energetically favorable configurations – Pt clusters for reducing conditions when single-atom traps on ceria are inefficient, and redispersed Pt single atoms for oxidizing conditions when single-atom traps provide strong Pt bonding.

KMC simulations within the generic model illustrate that oxidation and reduction treatments of metal deposits on single-atom catalyst substrates can be efficiently utilized in tailoring the morphology of both highly dispersed metal single-atom population and metal clusters on demand of a particular catalytic application. Understanding adatom and nanoparticle dynamics on single-atom catalyst substrates can also be utilized in future designs of sinter-resistant catalysts. We have demonstrated that the redispersion of metal deposit to single-atom traps can break the Ostwald ripening provided that the redispersion is allowed to be complete. We have also identified the conditions in which catalysts can be operated in a sinter-resistant manner in practical setups, e.g. in alternating oxidizing/reducing conditions.

**Acknowledgments**

FD, AT, VJ and JM acknowledge the support of the Czech Science Foundation, project No. GAČR 20-11688J. FD, MV, IK, TS and IM acknowledge Ministry of Education (LM2018116), the Czech Science Foundation (project No. 20-13573S) for financial support and the CERIC-ERIC Consortium for the access to experimental facilities. This project has received funding from the EU-H2020 research and innovation program (grant number 654360). NFFA-Europe provided access to the CNR-IOM theory facility within the framework of the NFFA Transnational Access Activity.

**Author Contributions**

FD, MV, IK, TS, IM, JM designed, performed and evaluated synchrotron experiments, FD, AT, VJ, JM designed, performed and evaluated laboratory STM and XPS experiments. MFC, SF designed, performed and evaluated DFT calculations, JM designed, performed and evaluated KMC simulations. SF, JM wrote the manuscript, all authors have read and commented on the manuscript.

**Methods:**

**Experimental procedures**

Experiments were performed at the Materials Science Beamline at the Elettra Synchtrotron in Trieste, Italy (XPS, SRPES, RPES), and in the laboratories of Department of Surface and Plasma Science at Charles University in Prague, Czech Republic (XPS, STM). Experiments were performed at UHV apparatuses using procedures described in detail in our previous publications [31], [32]. Particularly, $CeO_2(111)$ thin films were prepared by evaporation of Ce metal (Goodfellow, 99.95%) from Mo or Ta crucible on Cu(111) surfaces (MaTecK) in $5\times10^{-5}$ Pa $O_2$ (Linde, SIAD, 4.7). Density of monoatomic steps was controlled via deposition temperature, reduction/oxidation procedures or postdeposition annealing as summarized in Table S2. Density of monoatomic steps was determined from STM images via superimposing a manual outline of step edges on a properly scaled and oriented mesh of surface Ce atoms on $CeO_2(111)$, and counting the number of Ce atoms at step edges [31], [32]. Pt (Safina, Goodfellow, 99.95%) was deposited at 300 K, the amount of deposited Pt was controlled via a combination of quartz crystal microbalance (QCM) and XPS measurements of the attenuation of Ce 3d or Cu $2p_{3/2}$ signals [31], [32].

Upon Pt deposition, redispersion to $Pt^{2+}$ was performed by annealing in UHV at 700 K or lower as described in the text (Figure 1). Reduction treatments of the samples were performed via 1 min annealing the samples in $1\times10^{-5}$ Pa $CH_3OH$ at 600 K. Reduction treatments were repeated, eventually, until the ceria reduction indicated by RER (Figure S1 b, f, j) increased above 1.0. Oxidation treatments of the samples were performed via 10 min exposure to $1\times10^{-5}$ Pa $O_2$ at 300 K followed by flash annealing at 600 K in the same $O_2$ background. Oxidation treatments were repeated, eventually, until RER decreased below 0.1. After the oxidation the samples were flash annealed at 700 K in UHV.

In between the redispersion/reduction/oxidation steps oxidation state of Pt deposit and ceria support were characterized at 300 K by XPS, SRPES, and RPES. Overview of the evaluated data and the corresponding fitting and background subtraction is presented in Figures S1 and S2. STM experiments were performed at 300 K using a mechanically cut Pt-Ir STM tip. Complementary images of occupied states and empty states were obtained (Figure 2 b and Figure S3).

**DFT Calculations**

All calculations were based on the DFT employing ultrasoft pseudopotentials and the Perdew, Burke and Ernzerhof (PBE) generalised gradient-corrected approximation for the exchange and correlation functional [72], [73]. The electronic wave function and density were described with a plane-wave basis set employing energy cutoffs of 40 and 320 Ry, correspondingly.

Following the established practice we use the PBE+U approach to describe the electronic structure of ceria-based materials, employing the implementation of Cococcioni and de Gironcoli [74] which includes an additional Hubbard-U term to the Kohn–Sham functional that disfavors fractional occupancies of the Ce 4f states. The value of the parameter U was set to 4.5 eV [75]. All calculations have been performed using the Quantum-ESPRESSO computer package [76]. The minimum energy paths at 0 K have been computed using the CI-NEB algorithm using a Broyden scheme as implemented in the Quantum-ESPRESSO package. The reaction path was sampled with 8–12 intermediate images.

The adsorption of Pt atoms at ceria step edges was modelled with periodic supercells describing vicinal surfaces. Two of the lowest energy model steps proposed in Ref. [46] have been selected. Following this work, we label our two model step surfaces as Step Type I and Step Type II. In both models, the step edge exposes 3 independent O atoms. The step edge is separated by two (111) terraces and is oriented along the [110] direction. The lateral dimension for the supercells are 17.97 Å×11.67 Å and 15.72 Å×11.67 Å for Steps Type I and II, respectively, along the [112] and [110] directions. The supercell slabs comprised 9 atomic layers and their surfaces were separated by more than 11 Å of vacuum in the direction perpendicular to the (111) terrace. During the structural relaxation the lowest three atomic O–Ce–O layers were constrained to their bulk equilibrium coordinates, as well as the Ce atoms in the central O–Ce–O trilayers far from the step edge.

**KMC simulation**

We adapt a solid-on-solid (no vacancies, no overhangs) KMC model on a cubic lattice with a periodic boundary condition [51]. The model is implemented within the Bortz–Kalos–Lebowitz algorithm [77] allowing to set the simulation temperature and evaluate the simulation time. We consider mobile adatoms of the deposit (Pt) on substrates with fixed stepped geometry (ceria). Lattice parameters of the substrate and deposit are the same, i.e. no strain is considered. Pt–Pt interactions are approximated by a bond-counting scheme [50]. To account for 3D character of Pt nucleation, both lateral and vertical Pt–Pt bonds are considered. Geometry and bonding properties of monoatomic steps on the substrates in the simulation (Figure S6) are defined in a way that the interaction of Pt adatoms with adatom traps at lower step edges corresponds to the experimentally and theoretically determined interaction of Pt adatoms with monoatomic steps on $CeO_2(111)$ [31], [32]. Activation and binding energies in the model are summarized in Table 2. Binding energies in the model are scaled with respect to binding energies predicted by ab-initio calculations (Figure S8). This results in an effective shifting of the temperature scale of the simulation to lower values compared to experiment. Rescaling the activation energies in KMC allowed to obtain manageable computing times in the range of weeks for a 3600 s simulation (Figures 6, 7) on a single core.

Activation energy of Pt adatom hopping in the model is defined for 4 different situations: $E_a^{Pt \to Pt}$, $E_a^{Pt \to Ce}$, $E_a^{Ce \to Ce}$, and $E_a^{Ce \to Pt}$, depending on the type of the underlying atom to the Pt adatom before and after the hop. The 3D character of Pt growth is obtained by assigning the $E_a^{Pt \to Ce}$ hops one more broken bond – the vertical one. Activation energies for Pt hopping in the model are determined as follows: $E_a^{Pt \to Pt} = E_d + n E_n$, $E_a^{Pt \to Ce} = E_d + (n+1) E_n$, $E_a^{Ce \to Ce} = E_a^{Ce \to Pt} = E_d + n E_n + \delta_{i1} E_1^{ox} + \delta_{i2} E_1^{re}$, where $E_d$ is the activation energy for Pt adatom diffusion both on ceria and on Pt, $E_n$ is the energy associated with breaking a Pt-Pt nearest-neighbor bond, $n$ is the number of lateral Pt-Pt nearest neighbors before the hop ($n = 0…4$), and $i$ is the type of the step edge underlying the Pt adatom before the hop ($i = 0, 1, 2$ for none, oxidized and reduced step, respectively). Hopping frequencies are evaluated in a standard way, $v_x = v_0 \exp(-E_a^x/k_BT)$ where $v_0 = 10^{13}$ s$^{-1}$, $k_B$ is the Boltzmann constant and $E_a^x$ is the activation energy of an x-th process.

The charge state of Pt in the simulation, and the simulated Pt$^{2+}$ and Pt$^0$ SRPES signals are determined based on the local geometry of Pt atoms in the simulation according to previous experimental and theoretical findings [31], [32] as described in Figure S7. Experimental cycles of annealing in UHV, in reducing, and in oxidizing conditions are simulated by instantaneous switching of the binding energy of the step-edge traps in the model from deep to shallow or vice versa (cf. Figure S6). Evaluation of KMC configurations are performed at the end of particular simulation cycles without simulating the cooldown to 300 K and a subsequent heating up for the next cycle. Initial KMC configurations for the presented calculations are obtained in analogy to the experimental treatments by simulated deposition and annealing cycles. Parameters of the presented KMC simulations are summarized in Table S3.

Strong Metal-Support Bonding," *Science*, vol. 329, no. 5994, pp. 933–936, Aug. 2010, doi:10.1126/science.1191778.